\documentstyle[amssymb,prl,aps,epsfig,multicol]{revtex}

\begin{document}
\title{Finite size effect in the specific heat of ${\bf YBa_{2}Cu_3O_{7-%
\delta}}$}
\author{T. Schneider$^*$ and J. M. Singer}
\address{Physik-Institut, Universit\"at Z\"urich, Winterthurerstr. 190, CH-8057\\
Z\"urich, Switzerland}
\date{\today}
\maketitle

\begin{abstract}
We analyze high precision, high resolution microcalorimetric specific heat
data of detwinned ${\rm YBa_{2}Cu_{3}O_{7-\delta }}$ single crystals. It is
shown that a finite size effect must be taken into account when attempting
to extract critical properties. Contrary to previous claims, there is no
contradiction to a regular second order phase transition. Consistency with $%
3D$-$XY$ critical behavior is demonstrated.\newline

\noindent PACS numbers: 74.25.Bt, 64.60.Fr
\end{abstract}

\pacs{74.25.Bt,64.60.Fr}

\begin{multicols}{2}
\narrowtext
Recently, detailed high precision, high resolution microcalorimetric
specific heat data of detwinned ${\rm YBa_{2}Cu_{3}O_{7-\delta}}$ single
crystals were reported by Charalambous et al. \cite{Charalambous}.
Neglecting corrections to scaling and inhomogeneity induced finite size
effects it was argued that these measurements show definite
incompatibilities with $3D$-$XY$ critical behavior and provide evidence that
the specific heat critical exponents are asymmetric around $T_{c}$.
Furthermore it was concluded that the measurements are incompatible with a
regular second order phase transition. This claims contradict the hitherto
well established evidence, that in the experimentally accessible temperature
regime the critical properties of cuprate superconductors fall into the $3D$-%
$XY$ universality class \cite
{SchneiderAriosa,SchneiderKeller,Overend,Roulin,Hubbard,Kamal,Pasler,Torque,Hofer}%
.

In this letter we reanalyze the data of \cite{Charalambous} by taking an
inhomogeneity induced finite size effect into account. We identify a reduced
temperature regime bounded by $t_{\ast }^{+}\lesssim t^{+}\lesssim
t_{ph}^{+} $ and $t_{\ast }^{-}\lesssim t^{-}\lesssim t_{ph}^{-}$. As usual,
in the above expression the $\pm$-signs refer to $t>0$ and $t<0$,
respectively, where $t=(T-T_{c})/T_{c}$. Below $t_{\ast }^{\pm }$ a
microscopic inhomogeneity mediated finite size effect is identified. It
makes it impossible to enter the asymptotic scaling regime, while above $%
t_{ph}^{\pm } $ the temperature variation of the background, usually
attributed to phonons, becomes significant. To set the scale we note that in
the $\lambda $-transition of $^{4}$He there are no strains and the only
significant impurity is $^{3}$He, which occurs naturally only to the extent
of $1$ part in $10^{7}$, and which at that level has a negligible effect 
\cite{Ahlers}. Indeed, the critical properties can be probed down to $t^{\pm
}=10^{-9}$. In solids and, in particular, in cuprate superconductors such a
homogeneity can hardly be achieved. Even the best solid materials exhibit
phase transitions which are rounded over a range of $t^{\pm }\lesssim
10^{-4} $. For this reason finite size effects must be taken into account
when an attempt is made to extract critical properties from experimental
data. Our analysis clearly uncovers a finite size effect in the specific
heat data of detwinned ${\rm YBa_{2}Cu_{3}O_{7-\delta }}$ single crystals 
\cite{Charalambous}. On this basis it is shown that in an intermediate
temperature regime the data are fully consistent with $3D$-$XY$ critical
behavior and that the neglect of the finite size effect leads to erroneous
conclusions concerning the critical properties.

The letter is organized as follows: we sketch the scaling form of the
specific heat including the finite - size effect, as well as the corrections
to the leading behavior. On this basis we analyze the experimental data,
identify a finite size effect in terms of a scaling function and demonstrate
that in the experimentally accessible temperature regime around the
transition temperature $T_{c}$, limited by the finite size effect, there are
no incompatibilities with $3D$-$XY$ universality.

With the exception of the $\lambda $-transition in $^{4}$He it is in
practice very difficult to access the asymptotic critical regime $%
t\rightarrow 0$ \cite{Ahlers}. Due to inhomogeneities, a solid is
homogeneous over a finite length $L$ only. In this case, the actual
correlation length $\xi ^{\pm }(t)=\xi _{0}^{\pm }|t|^{-\nu }$ cannot grow
beyond $L$ as $t\rightarrow 0$, and the transition appears rounded. Because
of this finite size effect, the specific heat peak occurs at a temperature $%
T_{P}$ shifted by an amount $L^{-1/\nu }$ compared to the homogeneous
system. The magnitude of the peak located at temperature $T_{P}$ scales as $%
L^{\alpha /\nu }$ \cite{Privman}. For finite $L$ the specific heat
coefficient adopts the scaling form 
\begin{equation}
{C}/{T}={C}\left( t_{P},L=\infty \right) \ {+\ }L^{\alpha /\nu }{\cal F}%
_{1}(y),\quad y=tL^{1/\nu }.  \label{EQ1}
\end{equation}
where ${\cal F}_{1}(y)$ is a universal function of its argument and 
\begin{equation}
{C}(t_{P},L=\infty )=B^{+}(0,\infty )+A^{+}\left| t_{P}\right| ^{-\alpha }.
\label{EQ2}
\end{equation}
The study of finite size effects in liquid helium near the $\lambda $ point,
as the bulk is confined more and more tightly in one or more dimensions, has
been a very good ground for testing the finite size scaling theory \cite
{Schultka,Metha,Coleman}. 
From this work, the form of the scaling function ${\cal F}_{1}(y)$ is
reasonably well known for various confining dimensions and boundary
conditions.

In homogeneous systems and near the critical point, the general from of the
specific heat along path $t$ can be written as \cite{Domb} 
\begin{eqnarray}
C_{s} &=&-\frac{\partial ^{2}f_{s}}{\partial t^{2}}=\frac{A^{\pm }}{\alpha }%
|t|^{-\alpha }\left( 1+a_{c}^{\pm }|t|^{\Theta }+e_{c}^{\pm }t+\ldots
\right) +B^{\pm },  \nonumber \\
\pm &=&{\rm sign}(t).  \label{EQ3}
\end{eqnarray}
$f_{s}$ is the free energy density, $A^{\pm }$ and $\alpha $ are the leading
specific heat critical amplitude and exponent, respectively. $\Theta >0$
denotes the leading correction exponent, $a_{c}^{\pm }$, $e_{c}^{\pm }$ and $%
B^{\pm }$ are the amplitudes of the correction terms.\ These dimensionless
correction amplitudes are nonuniversal. They determine the size of the
critical region and will in general become larger near a crossover \cite
{Fisher}. Nevertheless, the leading behavior as $t\rightarrow 0$ is $%
|t|^{-\alpha }$, but even in the $\lambda $-transition of $^{4}$He this
corrections must be taken into account when attempting to extract reliable
estimates for the critical exponents and amplitudes.

The specific heat coefficient and $C_{s}$ are related by 
\begin{equation}
\frac{C}{T}=-\frac{\partial ^{2}F}{\partial T^{2}}=-V\frac{\partial ^{2}f_{s}%
}{\partial T^{2}}=-\frac{V}{T_{c}^{2}}\frac{\partial ^{2}f_{s}}{\partial
t^{2}}=\frac{VC_{s}}{T_{c}^{2}},  \label{EQ4}
\end{equation}
where $F$ is the free energy and $V$ the volume. When the hyperscaling
relation, $3\nu =2-\alpha $, holds, various critical amplitude combinations
adopt universal values \cite{SchneiderAriosa,Domb}. Examples include $%
a_{c}^{+}/a_{c}^{-}$ and 
\begin{equation}
\left( R^{\pm }\right) ^{3}=A^{\pm }\xi _{x,0}^{\pm }\xi _{y,0}^{\pm }\xi
_{z,0}^{\pm }.  \label{EQ5}
\end{equation}
$\xi _{i,0}^{\pm }$ is the critical amplitude of the correlation length
along $i=(x,y,z)$. In Tab. \ref{tab1} we list recent estimates for critical
exponents and amplitude combinations of the $3D$-$XY$ universality class.

To clarify the relevance of an inhomogeneity mediated finite size effect in
cuprate superconductors, we calculated the scaling function ${\cal F}_{1}(y)$
(see Eq. (\ref{EQ1})) from the reported data \cite{Charalambous}. The result
shown in Fig. \ref{fig1} exhibits the characteristic form of the scaling
function for a system in confined dimensions, i.e. corresponding to rod or
cube shaped inhomogeneities \cite{Schultka}. As a consequence, our analysis
clearly reveals that the ${\rm YBa_{2}Cu_{3}O_{7-\delta }}$ samples
considered here exhibit a finite size effect on a length scale $L$ ranging
from $300$ to $400\AA$ (Tab. \ref{tab2}). For this reason, deviations from $%
3D$-$XY$ critical behavior around $t_{p}$ (Tab. \ref{tab2}) do not signal
the failure of $3D$-$XY$ universality, as previously claimed \cite
{Charalambous}, but reflect a finite-size effect at work. Next we sketch our
data analysis and the derivation of the estimates listed in Tab. \ref{tab2}.

For this purpose we explore the occurrence of $\ 3D$-$XY$ critical behavior
in the intermediate temperature regime (i.e., away from $t_{P}$ ). Since $%
\alpha $ is small, $(A^{\pm }/\alpha )|t|^{-\alpha }\approx A^{\pm }/\alpha
-A^{\pm }\ln |t|$, it is advisable to plot the measured heat coefficient
data versus $\ln |t|$ or $\log _{10}|t|$. Such a plot is shown in Fig. \ref
{fig2} for sample YBCO3 \cite{Charalambous}. The upper one corresponds to $%
T<T_{c}$ and the lower one to $T>T_{c}$. The filled circles in Fig. \ref
{fig2} mark the intermediate scaling regime, while the open circles closer
to $T_{c}$ belong to the finite size affected regime. $T_{c}$ of the
fictitious homogenous system was adjusted to yield optimal consistency with
nearly straight branches in an intermediate regime, subjected to the finite
size constraint $T_{c}>T_{p}$. In this intermediate temperature region
(filled symbols), bounded by the region where the finite size effect
dominates and the regime where the phonon contribution leads to significant
corrections.\ Given this $T_{c}$, the remaining parameters $\widetilde{A}%
^{\pm }$ and $\widetilde{B}^{\pm }$ , entering the leading $3D$-$XY$ \ form 
\begin{equation}
{C}/{T}=\widetilde{A}^{\pm }10^{-\alpha \log _{10}|t|}+\widetilde{B}^{\pm },
\label{EQ7}
\end{equation}
were the fixed by a linear fit to the corresponding branches for $\alpha
=-0.013$ and $\widetilde{A}^{+}/\widetilde{A}^{-}=1.07$ (Tab. \ref{tab1}).
The resulting estimates are listed in Tab. \ref{tab2}. 

From Fig. \ref{fig2} it is seen that this procedure provides a reasonable
description and with that meaningful parameters needed to determine the
scaling function ${\cal F}_{1}(y)$, depicted in Fig. \ref{fig1}. Indeed, the
resulting critical amplitudes $A^{\pm }$ are consistent with the previous
estimate \cite{Roulin}, $A^{-}\approx 1.5mJ/(gK)$, for a ${\rm %
YBa_{2}Cu_{3}O_{7-\delta }}$ single crystal with $T_{c}\approx 93.05K$.
Further away from $T_{c}$ the deviations can be attributed to a temperature
dependent background contribution, accounting for nearly $95\%$ of the
specific heat coefficient. To illustrate this point we depicted in Fig. \ref
{fig3} the temperature dependence of the specific heat coefficient of sample
YBCO3 \cite{Charalambous} in a wider temperature range around $T_{c}$.
Hence, due to the finite size effect and the temperature dependence of the
background the intermediate regime is bounded by the temperature region
where the data depicted in Fig. \ref{fig2} fall nearly on straight lines.
Clearly, this interval is too small to estimate the corrections to scaling
(Eq. (\ref{EQ3})) reliably. In this context it should also be kept in mind
that any subtraction of the temperature dependent background interferes with
the corrections to scaling and that the essential limitation of the
intermediate regime arises from the finite size effect.

Given the estimates of the parameters listed in Tab. \ref{tab2} it is now
straightforward to construct the scaling function ${\cal F}_{1}(y)$ and to
calculate quantities of interest from the experimental data. Of particular
interest is the characteristic length of the inhomogeneities. Noting that at 
$T_{p}$ 
\begin{equation}
\left| t_{P}\right| ^{3}=\left( \frac{\xi _{x,0}^{+}\xi _{y,0}^{+}\xi
_{z,0}^{+}}{L^{3}}\right) ^{1/\nu }  \label{EQ8}
\end{equation}
holds, we can express $L$ with aid of the universal amplitude combination (%
\ref{EQ5}) in terms of $A^{+}$ and $t_{P}$: 
\begin{equation}
L=\frac{R^{+}}{\left( A^{+}\right) ^{1/3}|t_{P}|^{\nu }}.
\end{equation}
Interestingly enough, the resulting estimates, ranging from $300$ to $400\AA$
(Tab. \ref{tab2}) clearly point to rather microscopic inhomogeneities.

To summarize, we have shown that the specific heat measurements considered
here are remarkably consistent with $3D$-$XY$ critical behavior. An
inhomogeneity induced finite size effect makes it impossible, however, to
enter the asymptotic scaling regime, where the leading scaling term clearly
dominates. Nevertheless, the exposed consistency of the specific heat data
with $3D$-$XY$ critical behavior in the intermediate temperature regime,
combined with evidence emerging from penetration depth, other specific heat
and magnetic torque measurements, etc. \cite
{SchneiderAriosa,SchneiderKeller,Overend,Roulin,Hubbard,Kamal,Pasler,Torque,Hofer}%
, clearly reveals that the previous claim \cite{Charalambous}: `evidence for
asymmetric critical exponents, violating $3D$-$XY$ universality', based on
the identical measurements, must be attributed to an inappropriate data
analysis, neglecting finite size effects. Consequently, finite size effects
must be taken into account when attempting to extract critical properties
from the data \cite{Charalambous,Junod}. Finally, the resulting limitation
renders the experimental search for charged fixed point behavior \ in
cuprate superconductors \cite{Herbut,Calan} into an academic issue.

The authors are grateful to P. Gandit for providing the data files and to J.
Hofer, A. Junod and C. Meingast for very useful comments and suggestions on
the subject matter.

\bigskip\hrule\bigskip

{\ 
\begin{table}[tbp]
\begin{tabular}{|l|c|cc|}
\hline
& $^{4}$He \ \cite{Ahlers} & $\epsilon$-Exp. &  \\ \hline
$\alpha $ & $-0.013$ & -0.011$\pm 0.004$ & \cite{Guida} \\ 
$\nu $ &  & 0.6703$\pm 0.0015$ & \cite{Guida} \\ 
$\Theta =\omega \nu $ &  & $0.529\pm 0.009$ & \cite{Guida} \\ 
$A^{+}/A^{-}$ & $1.067\pm 0.003$ & $1.05$ & \cite{Domb} \\ 
$R^{+}$ &  & $0.36$ & \cite{Domb} \\ 
$R^{-}$ &  & $0.95$ & \cite{Domb} \\ 
$a_{c}^{+}/a_{c}^{-}$ & $1.03\pm 0.3$ & $1.17$ & \cite{Domb} \\ 
$a_{c}^{-}/a_{\rho _{s}}$ & $-0.068\pm 0.03$ & $-0.045$ & \cite{Domb} \\ 
\hline
\end{tabular}
\caption{Estimates of critical exponent and universal amplitude combinations
for the $3D$-$XY$ universality class.}
\label{tab1}
\end{table}
}

{\ 
\begin{table}[tbp]
\begin{tabular}{|c|c|c|c|}
\hline
& YBCO3 & UBC2 & UBC1 \\ \hline
$\widetilde{A}^{+}\ [mJ/(gK^{2})]$ & $-1.52$ & $-1.52$ & $-1.52$ \\ 
$A^{+}\ [mJ/(gK)]$ & $1.820$ & $1.824$ & $1.822$ \\ 
$A^{+}\ [10^{-20}cm^{3}]$ & $8.4$ & $8.4$ & $8.4$ \\ 
$\widetilde{B}^{+}\ [mJ/(gK^{2})]$ & $3.554$ & $3.562$ & $3.559$ \\ 
$\widetilde{B}^{-}\ [mJ/(gK^{2})]$ & $3.528$ & $3.544$ & $3.54$ \\ 
$T_{c}\ [K]$ & $92.12$ & $92.33$ & $92.29$ \\ 
$T_{P}\ [K]$ & $91.98$ & $92.24$ & $92.21$ \\ 
$t_{P}\ [10^{3}]$ & $-1.52$ & $-0.94$ & $-0.87$ \\ 
$L\ [\AA]$ & $290$ & $397$ & $419$ \\ \hline
\end{tabular}
\caption{Estimates for the critical amplitudes $\widetilde{A}^{+}$, the
background terms $\widetilde{B}^{\pm}$, the transition temperature $T_c$ of
the ficticious homogeneous system, the temperature $T_P$, where the specific
heat coefficient reaches its maximum value, and the characteristic length $L$
of the inhomogeneities. ($A^{+}=T_{c}\widetilde{A}^{+}\protect\alpha$, $%
A^{+}\ [cm^{3}] = \protect\rho /k_{B}10^{4}A^{+}\ [mJ/(gK)]$, $\protect\rho%
=6.37 g/cm^{3}$.)}
\label{tab2}
\end{table}
}

\end{multicols}

\bigskip\hrule

\newpage

{\narrowtext
\begin{figure}
\centerline{\epsfig{file=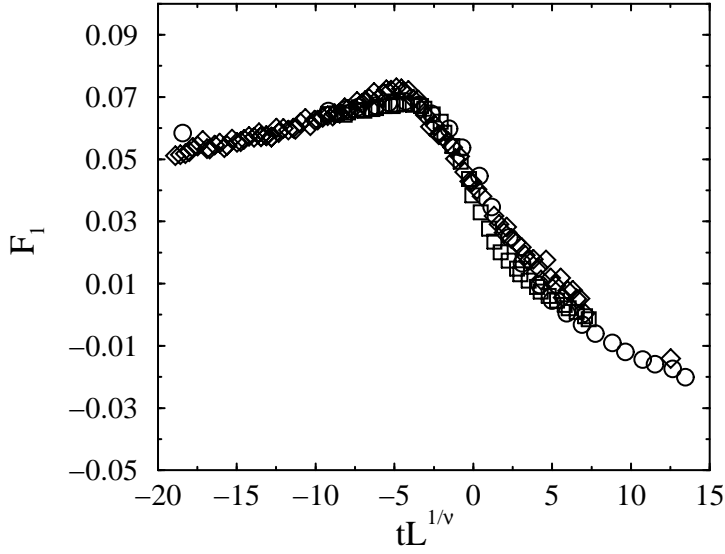,width=10cm,angle=0}}
\caption{Scaling function ${\cal F}_{1}(tL^{1/\nu})$ derived from the
experimental data of sample YBCO3 ($\diamond$), UBC2 ($\square$)
and UBC1 ($\circ$) with the aid of Eqs. (\ref{EQ1}) and
(\ref{EQ2}) using the parameters listed in Tab. \ref{tab2}.}
\label{fig1}
\end{figure}}

{\narrowtext
\begin{figure}
\centerline{\epsfig{file=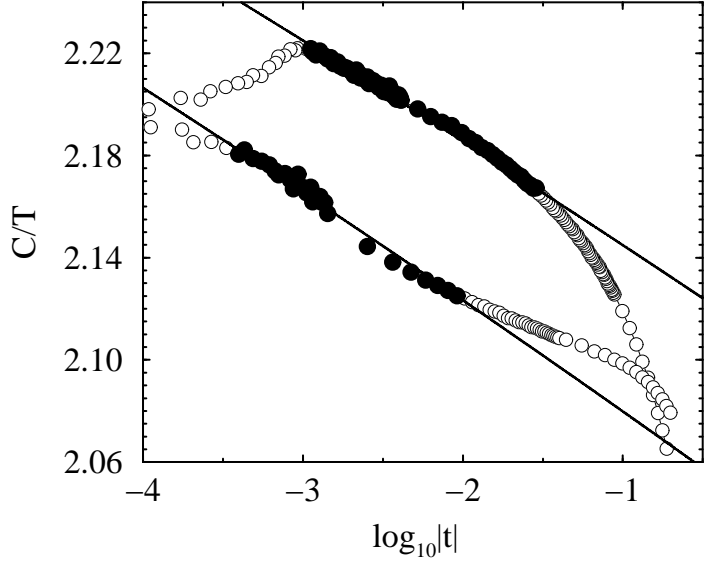,width=10cm,angle=0}}
\caption{Specific heat coefficient $C/T$ $[mJ/(gK^{2})]$ versus $%
\log_{10}|t| $ for ${\rm YBa_{2}Cu_{3}O_{7-\protect\delta}}$ (sample YBCO3)
with $T_{c}=92.12K$. The filled symbols denote the $3D$-$XY$ scaling regime,
whereas the open circles belong to the finite size and background affected
regimes, respectively. The solid lines are fits to Eq. (\ref{EQ7}) using the
filled data points only. }
\label{fig2}
\end{figure}
}

{\narrowtext
\begin{figure}
\centerline{\epsfig{file=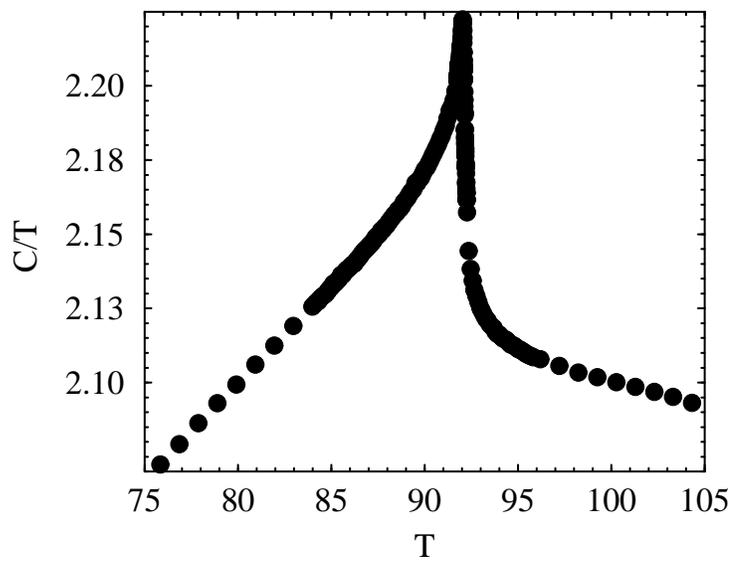,width=10cm,angle=0}}
\caption{Specific heat coefficient $C/T$ $[mJ/(gK^{2})]$ versus $T$ $[K]$ of
sample YBCO3 in a wider range around $T_{c}$. }
\label{fig3}
\end{figure}
}

\end{document}